\documentstyle[tighten,prl,aps,multicol]{revtex}
\begin{document}
\title{Characterizing GHZ Correlations in Nondegenerate
Parametric Oscillation via Phase Measurements}
\author{W.~J.~Munro and G.~J.~Milburn}
\address{Center for Laser Science,Department of Physics, University of
Queensland, QLD 4072, Brisbane, Australia}
\date{\today}
\maketitle
\begin{abstract}
\small
We present a potential realization of the Greenberger, Horne and
Zeilinger ``{\it all or nothing}'' contradiction of
quantum mechanics with local realism using phase measurement
techniques in a simple photon number triplet. Such a triplet could be
generated  using nondegenerate parametric oscillation.
\end{abstract}

\begin{multicols}{2}

Many of the traditional tests of quantum mechanics (using the Bell
Inequalities$^{\cite{Bell65,CHSH69}}$) have used parametric amplifiers or oscillators to
generate correlated photon number states$^{\cite{Burnham and Weinberg 
70,Friberg Hong and Mandel 85,Ghosh and Mandel 87}}$. When these correlated photon pairs
(the signal and idler) are passed through polarisers or 
beamsplitter/phase-shifters and, measured by
single photon detectors, a test of the Bell inequality
can be achieved. Such tests however require auxiliary conditions$^{\cite{CHSH69}}$ 
that lessen  (or call into doubt) the contradiction. Such contradictions
are very microscopic in nature as they involve single photon detection.
Multi-particle tests of the Bell inequality have also been proposed
using parametric amplification$^{\cite{Drummond 83,Munro and Reid 94}}$. 
No multi-particle test of the
Bell inequality has ever been experimentally considered.

The quantum states described by Greenberger, Horne and 
Zeilinger$^{\cite{GHZ 89,GHZ 90,Mermin2 90,Mermin3 90,Clifton 91}}$ (GHZ states)
give predictions contrary to those of all classical theories based on the
Einstein-Podolsky-Rosen$^{\cite{EPR35}}$ (EPR) premises of local realism.
The spin GHZ state is an entangled state of three spins specified by
stating that all spins are in the same direction. As this correlation can
be realized in two ways, the state is the sum of the two amplitudes
representing each way separately.  The resulting  interference between
these two amplitudes ensures that there is a particular result for triple
product spin measurements that can never occur. In contrast, a  classical
local hidden variable state exhibiting the same correlation (that is all
spins the same) will produce this forbidden result with a non-zero
probability. If this forbidden result were ever observed in a perfect
experiment, the quantum prediction would be incorrect. On the contrary,
never observing the forbidden result would verify quantum mechanics.
Unfortunately not observing an event is a difficult way to test a theory
experimentally.  Detector  inefficiencies may also lead to the non-observation 
of the forbidden result for reasons that have nothing to do
with quantum entanglement.

The GHZ paradox can be formulated as follows: Consider three spin $1\over
2$ particles in a state
$|\uparrow\rangle|\uparrow\rangle|\uparrow\rangle + |\downarrow\rangle|\downarrow\rangle|\downarrow\rangle$ where
the $\uparrow$ or $\downarrow$ specifies spin up or down along the
appropriate $z$ axis.
These particles originate in a spin conserving gendanken decay and
fly apart along three different straight lines in the $x-y$ plane.
Now because the spin vectors of distinct particles commute component
by component, we can simultaneously measure the $x$ component of one
particle  and the $y$ components of the remaining two. In fact for the 
given initial state the product of the results of the
three spin measurements $S_{x1} S_{y2} S_{y3}$, $S_{y1} S_{y2}
S_{x3}$, $S_{y1} S_{x2} S_{y3}$, where $S_{xi}$ and $S_{yi}$ represent the
spin along the horizontal and vertical directions, has to be $+1$ 
according to both quantum mechanics and local realism.
According to local realism the spin product $S_{x1} S_{x2} S_{x3}$
must also be unity. Such a product can also be calculated quantum
mechanically and in fact is found to be minus the product of all the 
three of them. To account for experimental situations where the spin product
predictions are not unity in size, Mermin$^{\cite{Mermin90}}$ derived the following
inequality based on local realism arguments
\begin{eqnarray}\label{oldinequality}
F=\left| S_{x_{1}} S_{x_{2}}\right. &S_{x_{3}}&-S_{y_{1}} S_{y_{2}}
S_{x_{3}} \nonumber \\
&-&\left. S_{y_{1}} S_{x_{2}} S_{y_{3}}-S_{x_{1}} S_{y_{2}}
S_{y_{3}}\right| \leq 2.
\end{eqnarray}

To date there have been no tests of the GHZ inequality given by (\ref{oldinequality}), due mostly to
the difficult nature of generating a triple spin state of the form
$|\uparrow\rangle|\uparrow\rangle|\uparrow\rangle + |\downarrow\rangle|\downarrow\rangle|\downarrow\rangle$. 
Recent developments by Laflamme {\it et. al}$^{\cite{Laflamme97}}$ have seen the generation
of a GHZ state using the proton and carbon
spins of trichloroethylene in NMR spectroscopy. They have shown
using state tomography techniques that a 95\% construction of the
triple spin state can be achieved. Because such an experiment was done using a molecule no significant
separation of the photon/carbon spins could be achieved and hence
a test of the locality condition implicate in the GHZ paradox could not be
made.

In this letter we propose a novel use of phase measurements to
test the GHZ correlations. We will show how a simple correlated photon
number triplet could be used to provide a definitive test. Such a state could be produced
via nondegenerate parametric oscillation where we have signal, idler
and pump modes. Discrete phase measurements are however difficult
to realize experimentally and hence we consider how a homodyne quadrature
phase amplitude measurement can provide a more realisable test.
In a homodyne measurement the signal field is coupled to  a strong local oscillator, hence providing very
efficient detection. Current homodyne detection efficiency$^{\cite{Polzik Carry and Kimble 92}}$ can exceed
$99\%$, thus providing a more stringent test by removing potential detection
efficiency loopholes$^{\cite{Kwiat et al 94,Freyberger et ak 95,Fry et al 96}}$. Also the  use of the strong local oscillator
field means that large intensities  are incident on the highly efficient
detectors.

A quantum entangled state shows correlations that cannot be explained
in terms of the correlations between local classical properties of 
the subsystems. In this letter we will describe a pure entangled state of three modes
in which the correlations are in photon number. More specifically,
the nature of the correlation can be succinctly stated by
saying that there are equal number of photons in each mode. As there
are many different ways to realize this fact, the total state is the sum
over amplitudes for all possible ways in which this correlation can be realized. This kind
of sum over amplitudes for correlations is characteristic of an entangled
state.

The question now arises as how best to see the quantum nature of the correlation.
Obviously it is not enough to measure photon number as this would not distinguish 
a mixed state with equal photon numbers in each mode, from the equivalent entangled 
pure state. In some sense we need to measure an observable which carries as little information as
possible about photon number in order to see the interference 
between all the possible ways in which the correlation in photon number can be realized. 
We conjecture that the best choice is the observable canonically conjugate 
to photon number: the canonical phase.

Pegg and Barnett$^{\cite{Pegg and Barnett 88,Pegg and Barnett 89}}$  have shown that a set 
of $s+1$ orthonormal phase states, with values of $\theta$ differing by $2 \pi /(s+1)$, can be 
generated from
\begin{eqnarray}\label{2}
| \theta_{\mu} \rangle = \exp \left[i \hat N \mu 2 \pi /(s+1) \right]
| \theta_{0} \rangle,\;\; \mu=0,\ldots,s
\end{eqnarray}
where $|\theta_{0} \rangle$ is the reference (or zero) phase state, 
$\hat N$ is the number operator and $\mu$ the particular
discrete phase we are interested in.  The values  for $\theta_{\mu}$ are
given by
\begin{eqnarray}
\theta_{\mu}=\theta_{0}+ {{ 2 \mu \pi}\over{(s+1) }}
\end{eqnarray}
which are spread evenly over the range $\theta_{0}\leq
\theta_{\mu}\leq\theta_{0}+ 2\pi$, where $\theta_{0}$ is the initial (or
reference) phase.

The probability of finding a generalized system $|\Psi\rangle$ in a particular phase state 
$| \theta_{\mu} \rangle$ is
\begin{eqnarray}
P_{\mu}(\theta_{0})=\left| \langle \Psi| \theta_{\mu} \rangle \right|^{2}
\end{eqnarray}
where $\mu$ labels the particular phase state, and $\theta_{0}$ is the
choice of initial phase.

We require large $s$ to describe an arbitrary phase for a general system. However, in the case of the
measurement schemes required for various quantum violations of classical inequalities such as
the Bell$^{\cite{Bell65}}$ and GHZ$^{\cite{GHZ 89,GHZ 90,Mermin2 
90,Mermin3 90}}$ (or Mermin higher spin$^{\cite{Mermin90}}$) inequalities,
all that is required and necessary is a binary result. Thus a discrete
phase measurement with $s=1$ suffices, that is two phase states are sufficient. 
If more phase states are chosen, for example $s=3$, a
binary result is still required for these particular quantum 
inequalities, which could be achieved by dividing or 
binning the phase states into two discrete distinct sets.
However this will not be ideal as to get this binary result we must
discard information. Such a process must lessen (or destroy) our potential GHZ
violation.

Production of a state of the form
\begin{eqnarray}
| \Psi \rangle ={1\over \sqrt{2}} | \uparrow \rangle |\uparrow \rangle  | \uparrow \rangle+
{1\over \sqrt{2}}  | \downarrow \rangle |\downarrow \rangle  | \downarrow \rangle
\end{eqnarray}
where $\uparrow,\downarrow$ represent the spin of the
particle, has been difficult to achieve experimentally. Reid
and Munro$^{\cite{Reid and Munro 92}}$ have considered previously a photon triplet state
\begin{eqnarray}
| \Psi \rangle = {1\over \sqrt{2}}  | 0 \rangle |0 \rangle  | 0 \rangle+{1\over \sqrt{2}}
| 1 \rangle |1 \rangle  | 1 \rangle
\end{eqnarray}
which can also be used to test the GHZ inequality. Production of this
triplet has yet to be realized. Potential for similar
photon triplet state production exists in parametric oscillation. The ideal nondegenerate
parametric oscillator may be  specified by an interaction Hamiltonian 
of the form
\begin{eqnarray}
H_{\rm int}= i\hbar \chi \left[ c^{\dagger} a b- c a^{\dagger} b^{\dagger}\right]
\end{eqnarray}
where $\hat a$, $\hat b$ and $\hat c$ are the boson annihilation operators
for the signal, idler, and pump modes, respectively and $\chi$ is the parametric
coupling constant. Initially preparing the pump mode in a single Fock 
state $|1\rangle$, with the signal and idler modes initially in vacuum 
states, it can be easily shown that 
\begin{eqnarray}\label{State}
| \Psi \rangle = c_{0} | 0 \rangle |0 \rangle  | 1 \rangle+c_{1} | 1 \rangle |1 \rangle  | 0 \rangle
\end{eqnarray}
can be generated where normalization requires $|c_{0}|^{2}+|c_{1}|^{2}=1$. 
The state (\ref{State}) is also a stable 
soliton solution$^{\cite{Drummond and Karen 98}}$ when the system is driven by a 
classical pump field coupled to mode $\hat c$.

Given the state (\ref{State}) one can calculate the probability of obtaining the phase states 
$\theta_{\mu_{1}},\theta_{\mu_{2}},\theta_{\mu_{3}}$ (where the 
labels $\mu_{1},\mu_{2},\mu_{3}$ corresponds to the $\hat a$, $\hat b$ 
and $\hat c$ modes respectively). For $s=1$, a choice of only two phase states, we have
\begin{eqnarray}
P_{\mu_{1}\mu_{2}\mu_{3}}&\;&\left(\theta_{0,1},\theta_{0,2},\theta_{0,3}\right)=
\left| \langle \Psi |
\theta_{\mu_{1}}\rangle|\theta_{\mu_{2}}\rangle|\theta_{\mu_{3}}\rangle\right|^{2}
\nonumber  \\
&=&{1\over 8}+{1\over 4} c_{0} c_{1} \cos \left[\left(\mu_{1}+\mu_{2}-
\mu_{3}\right) \pi+ \psi_{0}\right]
\end{eqnarray}
where $\mu_{i}$ is zero or one, and
$\psi_{0}=\theta_{0,1}+\theta_{0,2}-\theta_{0,3}$.
We explicitly note that our initial phases for the three particles $\theta_{0,i}$
can be expressed as one $\psi_{0}$.  To classify our binary result we say that
$\mu_{i}=1$ corresponds to a ``1'' measurement, while 
$\mu_{i}=0$ corresponds to a ``0'' measurement (for each of the
particles). If we consider a single particle, then there is a probability of
detecting it in the  ``1''  (labeling the probability
$P_{1}$) or the ``0'' state (labeling this  $P_{0}$).
Hence the probability of all three particles being in a ``1'' state is
\begin{eqnarray}
P_{111}\left(\psi_{0}\right)=
{1\over 8}-{1\over 4} c_{0} c_{1} \cos \left[\psi_{0}\right]
\end{eqnarray}
Similarly we can calculate the probability $P_{000}$
of all particles being in a ``0'' state
\begin{eqnarray}
P_{000}\left(\psi_{0}\right)=
{1\over 8}+{1\over 4} c_{0} c_{1} \cos \left[\psi_{0}\right]
\end{eqnarray}
Other probabilities such as $P_{0 0 1}$
can be calculated in an identical manner. It is necessary to point out
that probabilities such as $P_{0 0 1}$ and 
$P_{0 1 0 }$ are not identical due to the asymmetric 
initial state (\ref{State}).

We define the spin of a single particle $i$ as
\begin{eqnarray}
S_{i}(\theta_{0,i}) = P_{1}(\theta_{0,i})-P_{0}(\theta_{0,i})
\end{eqnarray}
where we are explicitly indicating that the spin depends on the
initial reference angle choice $\theta_{0,i}$. The spin product of the three particles
is then the product of each of the spins. Hence the triple spin product is
\begin{eqnarray}\label{triple}
S_{1} S_{2} S_{3} (\psi_{0})=- 2 c_{0} c_{1} \cos \left[ \psi_{0} \right]
\end{eqnarray}
where we use the label $S_{i}$ to represent the spin of the $i^{th}$
particle and the angle $\psi_{0}$ to represent the total simplified
initial phase choice. Given a triple spin product it is now possible to
examine the GHZ paradox.

Generally, previous authors$^{\cite{Mermin90,Reid and Munro 92}}$ have considered a GHZ inequality of the
form (\ref{oldinequality}). However because of our asymmetric initial
state, we will consider the
following inequality
\begin{eqnarray}\label{newinequality}
F=\left| S_{y_{1}} S_{y_{2}}\right. &S_{x_{3}}&-S_{x_{1}} S_{x_{2}}
S_{x_{3}} \nonumber \\
&-&\left. S_{y_{1}} S_{x_{2}} S_{y_{3}}-S_{x_{1}} S_{y_{2}}
S_{y_{3}}\right| \leq 2
\end{eqnarray}
which can be derived in an identical way to  (\ref{oldinequality}).
We note that according to local realism
\begin{eqnarray}
S_{y_{1}} S_{y_{2}}S_{x_{3}}=
S_{x_{1}} S_{x_{2}} S_{x_{3}} \times S_{y_{1}} S_{x_{2}} S_{y_{3}}\times
S_{x_{1}} S_{y_{2}} S_{y_{3}}
\end{eqnarray}
provided the magnitude of each spin product is one. Now according to local
realism, the triple spin
product $S_{y_{1}} S_{y_{2}}S_{x_{3}}$ has the same sign as the product
of the other three triple products. It can be shown that the three triple
spin products $S_{x_{1}} S_{x_{2}} S_{x_{3}}$, $S_{y_{1}} S_{x_{2}}
S_{y_{3}}$, $S_{x_{1}} S_{y_{2}} S_{y_{3}}$ all have the same negative
sign and hence $S_{y_{1}} S_{y_{2}}S_{x_{3}}$ should be negative.
Hence adding all four spin products together according to
(\ref{newinequality}) will give $F\leq 2$.

Next we need to relate these $S_{x}$ and $S_{y}$ to our $S(\theta_{0,i})$.
We specify that $S_{x}=S(0)$ and $S_{y}=S \left(\pi/2\right)$. It
can be easily shown using (\ref{triple}), our quantum mechanical triple spin product result, that
\begin{eqnarray}
F= 8 c_{0} c_{1}
\end{eqnarray}
Therefore $F>2$ if $c_{0} c_{1}>1/4$ and a violation is possible.
For the equal superposition in (\ref{State}) we have $c_{0}=c_{1}=1/\sqrt{2}$.
Therefore $F=4$ giving a maximal violation. If we have instead
used the inequality given by (\ref{oldinequality}), then $F= 4 c_{0}
c_{1}\leq 2$ for all $c_{0}$, $c_{1}$.

The scheme presented here requires a discrete phase measurement, which
has yet to be experimentally realized in the ultra high detector
efficiency limit. However, recent work by Gilchrist {\it et. al}$^{\cite{Gilchrist Deuar and Reid 98}}$, 
and Yurke and Stoler$^{\cite{Yurke and Stoler 97}}$ has suggested
how quadrature phase amplitude measurements may be used to test the Bell
inequality in the high detector efficiency limit. A homodyne based
scheme is considered next to provide a feasible phase measurement.

A quadrature phase-amplitude homodyne measurement $X(\theta)$ can achieved by combining
a signal field (say $\hat a$) with a local oscillator field (say
$\hat b$) to form  two new fields given by $\hat c_{\pm}=\left[ \hat a \pm
\hat b \exp \left( i \theta\right)\right] /\sqrt{2}$. Here $\theta$ is a
phase shift which allows the choice of particular observable to be
measured, for instance choosing $\theta$ as $0$ or $\pi/2$ allows the measurement 
of the conjugate phase variables $X(0)$ and $X(\pi/2)$ respectively. The homodyne
measurement is achieved by measuring using photodectors the intensities
of both the beams $c_{+}$ and $c_{-}$, and  then subtracting them to give
a photocurrent difference as $I_{d}=c_{+}^{\dagger} c_{+}-c_{-}^{\dagger}c_{-}$.
Using the definition for $c_{\pm}$ the photocurrent difference can be
rewritten in terms of the original signal and oscillator modes as
\begin{eqnarray}
I_{d}=\hat b^{\dagger} \hat a e^{-i \theta}+\hat b \hat a^{\dagger} e^{i\theta}.
\end{eqnarray}
In the limit of a large oscillator field we can make a
replacement of the $b$ mode by a real classical field $\epsilon$. Hence
\begin{eqnarray}
I_{d}=|\epsilon| \left(\hat a e^{-i \theta}+ \hat a^{\dagger} e^{-i
\theta}\right) =|\epsilon|  X(\theta)
\end{eqnarray}

Thus performing a measurement on the quadrature phase amplitude $X(\theta)$
yields a result $x(\theta)$ which ranges in size and sign.
For our state (\ref{State}), the probability of obtaining $x_{1}(\theta_{1}),
x_{2}(\theta_{2}), x_{3}(\theta_{3})$ (abbreviated as $x_{1}, x_{2},
x_{3}$)  for the three particles measured by individual homodyne
measurements is
\begin{eqnarray}
P_{x_{1}x_{2}x_{3}}\left(\psi_{0} \right)= 
\left| \langle x_{1}| \langle x_{2}| \langle x_{3}| \Psi\rangle \right|^{2}
\end{eqnarray}
For a given quadrature measurement $x_i$, we classify the result as ``1'' if
$x_i>0$ and ``0'' if $x_i<0$. The probability of obtaining the
result ``1'' for all three particles is then 
\begin{eqnarray}
P_{111}\left(\psi_{0}
\right)&=&\int_{0}^{\infty}\int_{0}^{\infty}
\int_{0}^{\infty} dx_{1}  dx_{2} dx_{3}
P_{x_{1}x_{2}x_{3}}\left(\psi_{0} \right) \nonumber \\
&=&{1\over 4}- {{ c_{0} c_{1}}\over 4} \sqrt{ \left({2 \over \pi} \right)^{3}}  \cos
\left[\psi_{0}\right]
\end{eqnarray}
Other probabilities such as $P_{0 0 1}$ can be
calculated in a similar fashion.

Defining the spin $S_{i}$ in terms of $P_{1}$ and $P_{0}$ as before, 
we can show that the triple spin product is given by
\begin{eqnarray}\label{triplespinproduct}
S_{1} S_{2} S_{3}(\psi_{0})=-2 c_{0} c_{1}
\sqrt{ \left({2 \over \pi} \right)^{3}}  \cos \left[ \psi_{0}\right]
\end{eqnarray}
and hence $F$ given by (\ref{newinequality}) reduces to
\begin{eqnarray}
F=8 c_{0} c_{1} \sqrt{ \left({2 \over \pi} \right)^{3}}
\end{eqnarray}
We maximize the discrepancy between quantum mechanics and local realism 
by choosing an equal superposition ($c_{0}
=c_{1}=1/\sqrt{2}$, and hence $F= 8 \sqrt{2}/  \sqrt{ \pi^{3}}\sim 2.0318>2$. 
Though this is a small violation, it is still a violation of the GHZ
inequality in the high detection efficiency limit.

A fundamental question that needs to be considered is why the magnitude of
the triple spin product in (\ref{triplespinproduct}) is not one as it
is in the discrete phase case (for the case $c_{0}=c_{1}=1/\sqrt{2}$)?
The answer is quite simple. Our homodyne measurement, while it may have
perfect detection efficiency, is not an accurate (or efficient) measurement of the
discrete phase.  This leads to a significant lessening of the size of the 
violation of the potential GHZ violation. The homodyne
measurement does however have its advantages. First and foremost, current
homodyne measurement technology allows detection efficiencies in excess
of $99\%$. Our model for homodyne assumes perfect efficiency detectors.
However, because of our small potential violation, the homodyne detection efficiency
would have to exceed $99.5\%$  in a real experiment provided
the initial state could be produced accurately. A second advantage is that as the homodyne
measurement involves a strong local oscillator via $I_{d}=\epsilon X(\theta)$ 
(with $\epsilon$ being the strength of the local oscillator),
the potential GHZ inequality violation could have a macroscopic nature.

To summarize, we have investigated  a triple photon
correlated state (that may be able to be produced by nondegenerate parametric
oscillation) that can be used to test the GHZ inequality proposed by Mermin. We have proposed how
discrete phase measurements could provide an effective test of the
inequality. In fact, a binary phase measurement could provide a
maximal violation of the GHZ inequality. As an approximation to the
binary phase measurement, we consider homodyne quadrature phase amplitude
measurements. Again a violation of the GHZ inequality is possible
although it is significantly reduced because it is an insensitive binary phase measurement. 
An advantage of the homodyne method however is that because it involves a strong local oscillator the
detection efficiencies are extremely high.

W.~J.~M would like to thank P.~D.~Drummond, M.~D.~Reid, S.~Schneider and 
M.~J.~Gagen for their valuable discussions and assistance.

\end{multicols}

\end{document}